\documentclass[prl,aps,singlecolumn,superscriptaddress,preprintnumbers,nopacs,floatfix,amsmath,amssymb]{revtex4}

\usepackage{graphicx}
\usepackage{dcolumn}
\usepackage{bm}

\usepackage{amsmath}
\usepackage{graphicx}
\usepackage{amssymb}
\usepackage{mathrsfs}
\usepackage{bbm}
\usepackage{epsfig}

\newcommand{\be}{\begin{eqnarray}}
\newcommand{\ee}{\end{eqnarray}}

\begin{document}

%
%
%
\title{ Discrete Nonlinear Schr\"odinger Equation, Solitons  \\ and Organizing Principles for  Protein  Folding }

\author{Nora Molkenthin}
\affiliation{
Laboratoire de Mathematiques et Physique Theorique
CNRS UMR 6083, F\'ed\'eration Denis Poisson, Universit\'e de Tours,
Parc de Grandmont, F37200, Tours, France}
\affiliation{Department of Physics and Astronomy, Uppsala University,
P.O. Box 803, S-75108, Uppsala, Sweden}
\author{Shuangwei Hu}
\affiliation{
Laboratoire de Mathematiques et Physique Theorique
CNRS UMR 6083, F\'ed\'eration Denis Poisson, Universit\'e de Tours,
Parc de Grandmont, F37200, Tours, France}
\affiliation{Department of Physics and Astronomy, Uppsala University,
P.O. Box 803, S-75108, Uppsala, Sweden}
\author{Antti J. Niemi}
\affiliation{
Laboratoire de Mathematiques et Physique Theorique
CNRS UMR 6083, F\'ed\'eration Denis Poisson, Universit\'e de Tours,
Parc de Grandmont, F37200, Tours, France}
\affiliation{Department of Physics and Astronomy, Uppsala University,
P.O. Box 803, S-75108, Uppsala, Sweden}

\begin{abstract}
\noindent
We introduce a novel
generalization  of the discrete nonlinear Schr\"odinger equation. It
supports solitons that describe how proteins  fold. As an example 
we scrutinize the villin headpiece HP35, an archetypal protein for testing both 
experimental and theoretical approaches to protein folding. 
Using explicit soliton profiles we construct its carbon backbone 
with an unprecedented accuracy. 
\end{abstract}




\maketitle

The discrete nonlinear Schr\"odinger equation \cite{dnls}  is a prime example of a universal nonlinear equation.
The equation originally  appeared in connection of a study of polarons in  molecular crystals \cite{hol}.  
It supports both stationary and time dependent soliton solutions that were first  introduced  
to describe Davydov solitons in proteins \cite{scott},  then found applications 
to  the crystalline state of acetanilide  \cite{eil}, and subsequently 
emerged in the study of optical waveguides and Bose-Einstein condensates \cite{lin}.
Today the discrete nonlinear Schr\"odinger equation together with its generalizations (GDNLS) 
form a very actively studied family of nonlinear equations that  are widely employed to describe a multitude of phenomena  
in  disparate  physical, chemical and biological scenarios \cite{dnls}-\cite{scottb}. 

Here we  introduce a novel  generalization of the discrete nonlinear Sch\"odinger equation that
governs the organizing principle for  protein folding \cite{dill},  arguably among the most important 
unresolved phenomena in modern science. Our version of the GDNLS equation stems from a discrete
lattice model introduced in \cite{ulf} to describe the statistical properties of folded chiral homopolymers.  
A recent Monte Carlo investigation \cite{chernodub} has suggested that this model might support 
soliton-like solutions, and furthermore that these solitons might accurately model the folded 
protein structures that are stored in the Protein Data  Bank (PDB) \cite{pdb}.  The goal of the present article is 
 to adapt and  develop the powerful exact and numerical techniques of GDNLS 
 equations to address and resolve the organizing principles that underlie protein folding,  whereupon 
 a folded  protein becomes very accurately described
by a set of heteroclinic standing wave solutions {\it i.e.} dark solitons of an appropriate GDNSL equation.

Our GDNLS equation for protein folding  originates from
the following energy functional \cite{ulf}, \cite{chernodub},
\[
E = - \sum\limits_{i=1}^{N-1}  2\, \kappa_{i+1} \kappa_i  + \sum\limits_{i=1}^N
\left\{ 2 \kappa_i^2 + c\cdot (\kappa_i^2 - m^2)^2\right\}
\]
\begin{equation}
+ \sum\limits_{i=1}^N \left\{ b \, \kappa_i^2 \tau_i^2 + d \, \tau_i + e \, \tau^2_i +
q \, \kappa^2_i \tau_i
\right\}
\label{E}
\end{equation}
We take $\kappa_i$ to be periodic, $\kappa_i \in [-\pi, \pi] \   {\rm mod}(2\pi)$. It
is our primary variable and  subject to both local and nearest-neighbor interactions. In applications to
protein folding we identify $\kappa_i$  with the discrete signed Frenet curvature of the protein backbone, at the position of the $i^{th}$ 
$C_\alpha$ carbon. 
The variable $\tau_i \in [-\pi,\pi] \   {\rm mod}(2\pi)$ is  a periodic auxiliary variable and 
only subject to local interactions, it describes  the discrete Frenet torsion at the site $i$ of the protein backbone.
Finally, $(b,c,d,e,m,q)$ are  {\it global} parameters, they are specific to a given secondary superstructure. 

Our GDNLS equation emerges as follows:  We first eliminate the auxiliary variable by varying the energy functional with respect to
$\tau_i$. This  gives us an equation of motion to resolve for  $\tau_i$ in terms of $\kappa_i$,
\[
\frac{\partial E}{\partial\tau_i} = 2b\kappa_i^2 \tau_i + 2 e \tau_i + d + 
q\kappa^2_i \  = 0 
\]
\begin{equation}
\Rightarrow \ \ \tau_i [\kappa_i] = - \frac{1}{2} \frac{d + q\kappa_i^2}{e + b\kappa_i^2}
\label{Etau}
\end{equation}
We then perform a variation of the energy functional with respect to $\kappa_i$,  and  
substitute $\tau_i[\kappa_i]$ from  (\ref{Etau}) into the ensuing 
equation of motion to arrive at our GNLS equation
\begin{equation}
\kappa_{i+1} - 2 \kappa_i + \kappa_{i-1} \ = \ U' [\kappa_i] \kappa_i  \ \equiv\ \frac{dU[\kappa_i]}{d\kappa_i^2} \ \kappa_i \ \ \ \ (i=1,...,N)
\label{Ekappa}
\end{equation}
(with $\kappa_{0} = \kappa_{N+1} = 0$.)
This equation determines the stationary points of the following GDNLS Hamiltonian
\[
H \ = \  - 2\sum\limits_{i=1}^{N-1} \kappa_{i+1} \kappa_i \ + \ \sum\limits_{i=1}^N  \left\{ 2 \kappa_i^2 + U[\kappa_i] \right\}
\]
where the potential has the following functional form
\[
U[\kappa] = -  \left( \frac{ bd - eq}{2b}\right)^2 \cdot  \frac{1}{e+b\kappa^2}  - 
\left( \frac{q^2 + 8bcm^2}{4b}\right) \cdot  \kappa^2 + 
c\cdot \kappa^4
\]
Here the second and the third term are familiar in the context of the nonlinear Schr\"odinger equation \cite{dnls}-\cite{scottb}. But the 
first term appears to be novel in the present context, it resembles  the potential term for the relative coordinate 
in the two-body Calogero  model \cite{galo}.

If we properly  choose the parameters in (\ref{E}) so that the potential $U[\kappa]$ has two separate local minima, we can 
easily extend the results in \cite{herr} to ensure the existence of a dark soliton  solution that interpolates between these two minima. 
Such a qualitative form of $U[\kappa]$ typically follows if away from the vicinity of $\kappa = 0$ the potential  
becomes dominated by the second contribution to $E$ in (\ref{E}).  This is the 
familiar double-well potential term, with minima at $\kappa = \pm m$. It turns out that in applications to protein
folding the parameters should  indeed be chosen in this manner and
a dark soliton is a configuration that interpolates from the ground state in the vicinity of $\kappa_1 \approx \pm m$ 
to the ground state in the vicinity of $\kappa_N \approx \mp m$, as we traverse the backbone.  
When we compute $\kappa_i$ from (\ref{Ekappa}) and  $\tau_i$ from (\ref{Etau})  and integrate the ensuing discrete Frenet 
equation we obtain a  $N$-vertex polygonal chain such that a ground state  with  $\kappa \approx \pm m$ 
and $\tau$ given by (\ref{Etau})  is a helix, with the dark soliton  describing a loop that connects two helices.

We follow \cite{herr} to solve (\ref{Ekappa})  iteratively by locating a fixed point of 
\begin{equation}
\kappa_i^{(n+1)} \! =  \kappa_i^{(n)} \! - \epsilon \left\{  \kappa_i^{(n)} U'[\kappa_i^{(n)}]  
- (\kappa^{(n)}_{i+1} - 2\kappa^{(n)}_i + \kappa^{(n)}_{i-1})\right\}
\label{ite}
\end{equation}
Here  $\{\kappa_i^{(n)}\}_{i\in N}$ denotes the $n^{th}$ iteration of an initial configuration  $\{\kappa_i^{(0)}\}_{i\in N}$ and $\epsilon$ is some 
sufficiently small but otherwise arbitrary numerical constant. It is obvious  that a fixed point of (\ref{ite}) satisfies the GDNLS equation (\ref{Ekappa}).

In our simulations we start from an initial configuration  $\{\kappa_i^{(0)}\}_{i\in N}$  chosen to have the same overall
topology as the desired dark multi-soliton solution. We take $\kappa_i^{(0)}$ to have the profile of a piecewise constant 
step-function, the  constant values  approximate the true potential minimum. They correspond to the $\alpha$-helices and $\beta$-strands 
in the protein backbone. 
There is a step with a change of sign in $\kappa_i^{(0)}$ at each lattice site $ i= N_a$ where a backbone loop
is centered.  Notice that as it stands, the energy functional (\ref{E}) has 
the $\kappa \leftrightarrow -\kappa$ reflection symmetry that may not be exactly realized by the desired dark soliton profiles - the $\alpha$ helices are not
ideal, and there are proteins where a loop connects an $\alpha$-helix with a $\beta$-sheet. 
Thus we explicitely break this symmetry using the parameter $m$, and for this we set
\[
m \to  m_a \ \ \ \ {\rm for} \ N_{a-1} \leq i \leq N_a 
\]
along the chain.  Typical values  for $m_a$ are 
$m_a \approx \pm \pi/2$ for $\alpha$-helix, and $m_a \approx \pm 1$ for $\beta$-strand.

We have performed extensive numerical investigations of the  dark soliton solutions to (\ref{ite}). We have 
found that for proper values of the parameters these solitons can be combined into multi-solitons that  together with (\ref{Etau})
give a {\it very} high accuracy approximation of various folded protein structures that are stored in the Protein Data Bank \cite{pdb}, with the
 $\alpha$-helices and $\beta$-strands as the ground states and interpolated by dark solitons that describe the protein loops.

As an example we here scrutinize the dark two-soliton that models  the chicken villin headpiece 
subdomain HP35 (PDB code 1YRF), a naturally existing 35-residue protein that has three $\alpha$-helices 
separated from each other by two loops. The structure of HP35 is very  robust and since the protein is also a very  fast folder,  
the folding time is around 4$\mu$s, together with the engineered version (2F4K in PDB) and the 
very similar HP36 (1VII in PDB), the HP35 has become the subject to very extensive 
studies both experimentally \cite{knight}-\cite{wick} and theoretically \cite{pande}-\cite{fred}. Indeed, HP35  is now
a paradigm platform  for testing approaches to protein folding. 

According to \cite{meng}, the root mean square distance (RMSD) between the NMR spectroscopy and the x-ray crystallography
structures of HP35 is around 1.3 \.A for the $C_\alpha$ carbons.  The overall resolution of the presumably 
more accurate x-ray data is 1.07\.A \cite{chiu}. 
 
The authors of \cite{pande}-\cite{fred} report on the construction of native and near-native folds using various methods and 
with both explicit and implicit water. For example the proposed native fold in \cite{duan} deviates in average 
around 1.63 \.A in $C_\alpha$ RMSD from  the x-ray data \cite{chiu} for the sites 2-34 (counting from the N-terminus). 
The article also describes a single trajectory that reaches  
a value of 0.39 \.A  in RMSD  {\it i.e.} a distance that is about half the radius of 
a single carbon atom $[sic]$.  The authors of \cite{fred} report very similar results, with a proposed native fold average $C_\alpha$ RMSD 
around 1.54 - 1.65 \.A for the sites 2-34. They also report on a single trajectory that reaches $C_\alpha$ 
RMSD value 0.55 \.A. 

We shall now explain how the dark solitons of (\ref{ite}) quite effortlessly enable us to construct a
backbone  with 0.74 \.A RMSD accuracy for the $C_\alpha$ carbons, for the sites 3-33 (counting from the N-terminus);
The reason we do not consider the entire chain is
that in order to compute the local curvature from the three dimensional space coordinates
we need to know the coordinates of three adjacent $C_\alpha$ carbons, and for the computation 
of local torsion we need four.  

We convert  the PDB data for the $C_\alpha$ carbons to 
the local curvature  and torsion. 
The result is shown in Figure 1.
\begin{figure}[h]
        \centering
                \includegraphics[width=0.8\textwidth]{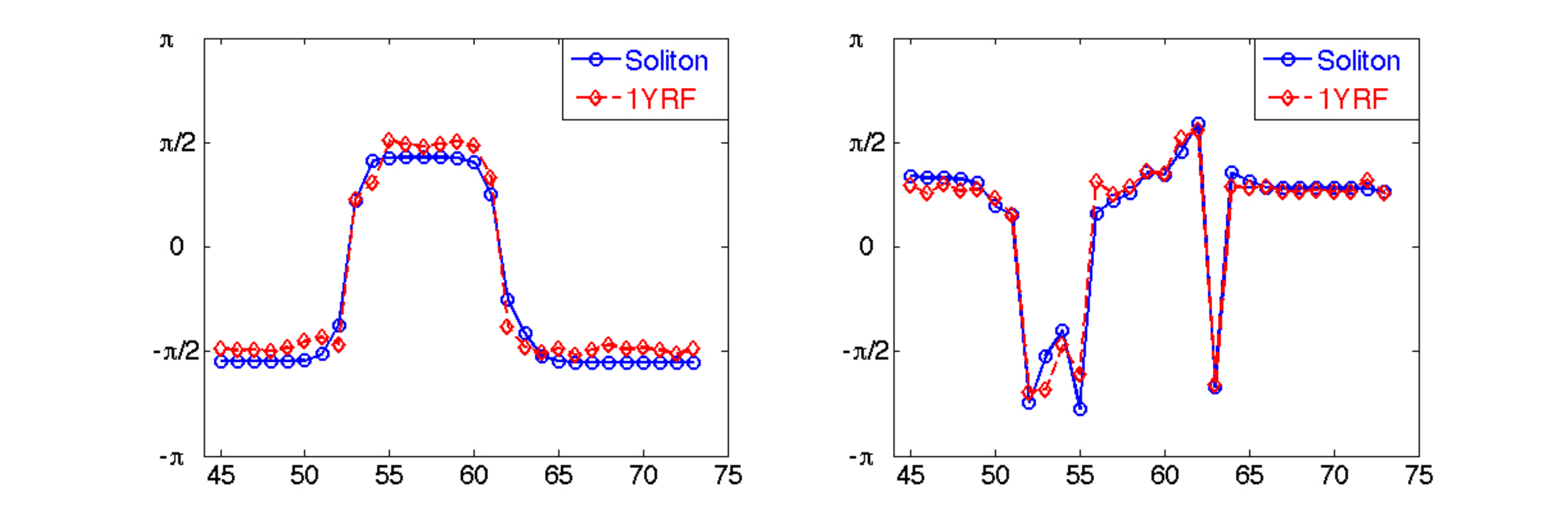}
        \caption{{ \it
      {\tt (Left):} The  bond angles $\kappa_i$ of 1YFR (red) for the sites 3-33 (45-78 in the PDB indexing convention) and their approximation by a soliton
       solution to equation (\ref{Ekappa}) (blue).
       {\tt  (Right):} The torsion angles $\tau_i$ of 1YRF (red) for the sites 3-33 (45-78 in the PDB indexing convention) and their approximation by a soliton
       solution to equation (\ref{Etau}) (blue).
                }}
       \label{Figure 1}
\end{figure}
From the $\kappa_i$ profile we conclude that the $C_\alpha$ backbone of 1YRF 
consists of two dark solitons. These correspond to the two loops of 1YRF and  are  located around
the sites 49-53 (PDB indexing) and 58-62 in Figure 1, respectively.  These solitons  interpolate between ground states that correspond to
the three $\alpha$-helices of 1YRF. The first helix  is located between the sites 42-49, the second  between the loops around
sites 53-58  and 
the third occupies the remaining sites starting from 62 in Figure 1.  While the two soliton profiles $\{\kappa_i\}$ are  
clearly identifiable, the profile of $\{\tau_i\}$ 
is substantially less regular and  {\it a priori} one may expect that the strong  irregularity  in  $\{\tau_i\}$  
reflects the amino acid differences in the side chains. However, we find 
that this is {\it not} the case. The $\{ \tau_i\}$ profile can be computed {\it very} accurately from (\ref{Etau}) in terms of the 
soliton profile $\kappa_i$, the apparent irregularity reflects {\it solely} the mod($2\pi$) multivalued 
character of a periodic variable.

To construct the soliton profile we introduce for each of the two would-be solitons 
the parameters $(b,c,d,e,m,q)$: There is one set of parameters for the 
sites $i$=3-13 (counting from $N$ terminus) and another set of parameters
for the remaining sites. 
We construct the ensuing soliton solution of (\ref{Ekappa}) by iterating (\ref{ite}) to a fixed point, 
starting from the  initial profile which is a step-function located at the solitons.  
We compute the RMSD  between the fixed point and 1YRF.
We then change the parameters randomly and compute the  new soliton profile, always starting from the
same piecewise constant initial profile for the $\kappa^{(0)}_i$. We compare its 
RMSD  to 1YRF with that obtained for the first set of initial parameters using the standard  Metropolis 
algorithm deviced to minimize RMSD. By repeating these steps
in combination with simulated annealing we eventually produce our final soliton solution.  The construction of a folded structure
takes about 10 hours using a single processor in a MacPro desktop computer

Figure 2 compares our minimal RMSD two-soliton configuration with the 1YRF backbone constructed 
from the x-ray data, for the sites $i$=3-33. 
\begin{figure}[h]
        \centering
                \includegraphics[width=0.45\textwidth]{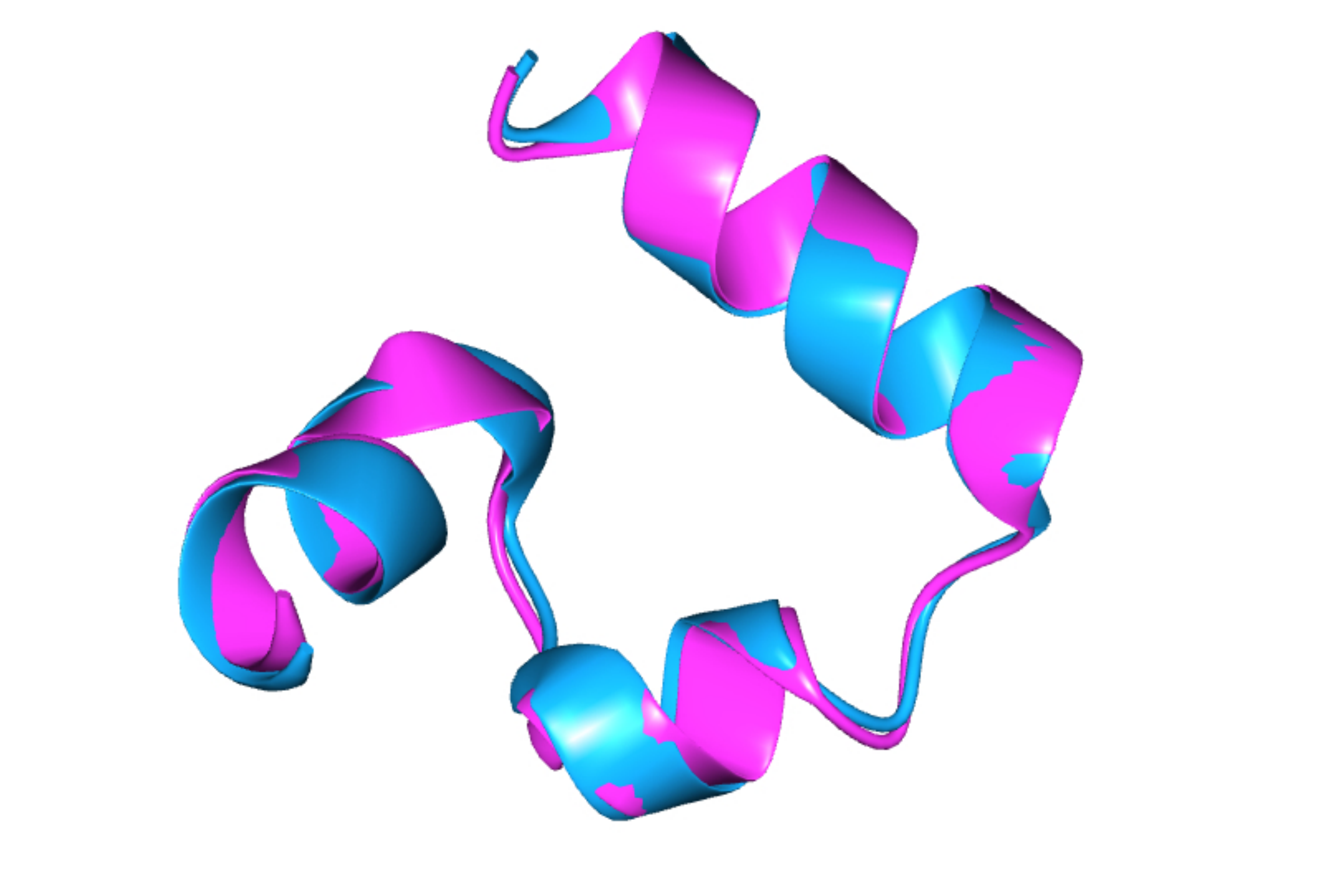}
        \caption{{ \it
      {Comparison between 1YRF backbone (red) and a soliton solution of (\ref{E}) (blue). The RMSD distance
      is 0.74 \.A. }
      }}
        \label{Figure 2}
\end{figure}
The RMSD  between the two configurations is 0.74 \.A, well below the overall resolution
of the x-ray data (which is 1.07 \.A). Consequently our dark soliton pair  describes the native 1YRF 
backbone  for all practical purposes, and with an accuracy comparable to that of 
the radius 0.70 \.A of a carbon atom. In Table 1 we provide the parameter values for this 
configuration, together with the parameter values for the best individual solitons we have found for the two loops. 
It is visible from the data that for values of $\kappa$ away from $\kappa \approx 0$ the potential energy is 
indeed strongly dominated by the double well contribution {\it i.e.} second term in (\ref{E}), as we have expected.
\begin{table*}[htbp]
	\centering
\begin{tabular}{|c|c|c|c|c|c|c|c|}
\hline  parameter & $b$ & $c$ & $d$ & $e$ & $q$ & $m_1$ & $m_2$   \tabularnewline 
\hline \hline
$1^{st}$ set & -0.000646646 & 0.227432 & 0.0141014 & 0.00162415& -0.0051673 & 1.68028 & 1.68844   \tabularnewline
\hline 
$2^{nd}$ set  & -0.0001126726 & 0.418995 & 0.000670547 & 0.00025209 & -0.000318858 & 1.69553 & 1.53529  \tabularnewline
\hline 
soliton-1& -0.000516175 & 0.662187 & 0.0081804 & 0.00110988  & -0.00356352  & 1.48643 & 1.48167  \tabularnewline
\hline 
soliton-2  & -0.0000443408 & 0.577717 & 0.000294502 & 0.0000936295 & -0.000132267 & 1.53816 & 1.54597 \tabularnewline
\hline 
\end{tabular}
\caption{The parameter values for the two-soliton solution that describes the entire 1YRF protein with accuracy 0.74 \.A,
for its  first  ($1^{st}$) loop (sites 2-13) and second ($2^{nd}$) loop (sites 14-33). We also present the parameter 
values for a dark soliton (soliton-1) that describes the first  loop with accuracy 0.76 \.A, and the
corresponding values for a dark soliton (soliton-2) that describes the second loop with accuracy 0.58 \.A.}
	\label{tab:para}
\end{table*}

\vskip 0.2cm
We have found that folded proteins can be described by dark soliton solutions of
a generalized discrete nonlinear Schr\"odinger equation. This equation involves  only {\it global } 
parameters specific to a secondary superstructure,
and the final
protein configuration is 
determined by a {\it single} function. In the
particular case of 1YRF where there are several high precision results to compare with,
we have constructed a two-soliton configuration that describes the native backbone with an
atomary level accuracy which is around one \.Angstr\"om  {\it less} than the present consensus 
value obtained in molecular dynamics simulations.  Among our future challenges is the enumeration and
modeling of the different secondary superstructures in PDB and to  develop a soliton basis for protein structure prediction.
Indeed, we find it  remarkable that in our construction we assume {\it
nothing} about the 
{\it details} of the amino acid sequence, we only describe a homogeneous $C_\alpha$ backbone.  Thus  
it is very unlikely that the common point of view  that folding is  {\it mainly} driven by side-chain interactions can be the full explanation.
Instead, our results suggest the presence of a {\it strong} contribution from
backbone hydrogen bonding  \cite{rose}, \cite{baldin}. The detailed amino acid structure then
breaks the translation invariance along an otherwise homogeneous chain, and amino acids in particular
 structural disruptor proline determine the location and the size of the 
 loops {\it a.k.a.} dark solitons. In this manner  the  folding geometry {\it is } dictated by genome.

\vskip 0.2cm
Our research is supported by grant from the Swedish Research Council (VR). N.M. thanks M. Herrmann for communications. We all
thank Martin Lundgren for discussions. 

\vfill\eject


\end{document}